\documentclass[12pt]{article}
\usepackage{graphicx,bm,epsf,amsmath,amsthm,amscd,amsfonts,amssymb,amsbsy,color}
\usepackage{type1ec}
\usepackage[T1,T2A]{fontenc}      
\usepackage[cp1251]{inputenc}   
\providecommand{\href}[2]{#2}
\usepackage[hyphens,spaces,obeyspaces]{url}
\usepackage[colorlinks,breaklinks,allcolors=blue]{hyperref}
\usepackage[hdvips]{hypdvips} 
\usepackage{doi}
\newcommand{\e}{{\rm e}}
\newcommand{\m}{q}

\newcommand{\be}{\begin{equation}}
\newcommand{\ee}{\end{equation}}
\newcommand{\nn}{\nonumber}
\newcommand{\noi}{\noindent}
\newcommand {\mm}[1]{\quad\mbox{#1}\quad}
\newcommand {\MM}[1]{\qquad\mbox{#1}\qquad}
\newcommand{\bc}{\begin{center}}
\newcommand{\ec}{\end{center}}
\newcommand{\bi}{\begin{itemize}}
\newcommand{\ei}{\end{itemize}}
\begin{document}
\title{\bf A new special function related to a discrete Gauss-Poisson distribution
and some physics of the cell model with Curie-Weiss interactions\footnote{Preprint
\href{https://arxiv.org/abs/2412.05428}{ArXiv {\tt 2412.05428} [math.CA] (2024)}
with corrections of several misprints}
}
\author{\bf O. A. Dobush and M. A. Shpot\\
{\it Institute for Condensed Matter Physics, 79011 Lviv, Ukraine}
}
\date{\today}
\maketitle
\begin{abstract}
\noindent
Inspired by previous studies in statistical physics $\big[$see, in particular,
Kozitsky at al., \emph{A phase transition in a Curie-Weiss system with binary interactions}, Condens. Matter Phys. {\bf 23}, 23502 (2020)$\big]$
we introduce a discrete Gauss-Poisson probability distribution function
\begin{equation}\label{GPD}\tag{A1}
p_{GP}(n ;z,r)=\left[R(r;z)\right]^{-1}\frac{{\rm e}^{zn}}{n!}\,\mbox{e}^{-\frac 12\,rn^2}
\end{equation}
with support on $\mathbb N_0$ and parameters $z\in\mathbb R$ and $r\in\mathbb R_+$. The probability mass function $p_{GP}(n ;z,r)$ is normalized by the special function $R(r;z)$, given by the infinite sum
\begin{equation}\label{R}\tag{A2}
R(r;z)=\sum_{n=0}^\infty\frac{{\rm e}^{zn}}{n!}\,{\rm e}^{-\frac 12\,rn^2},
\end{equation}
possessing extremely intersting mathematical properties.
We present an asymptotic estimate $R^{(\rm as)}(r;z\gg1)$ for the function $R(r;z)$
with large arguments $z$, along with similar formulas for its logarithm and
logarithmic derivative.
These functions exhibit very interesting oscillatory behavior around their asymptotics, for parameters $r$ above some threshold value $r^*$.
Some implications of our findings are discussed in the context of the
Curie-Weiss cell model of simple fluids.
\end{abstract}



\section{Introduction}

Although a number of physical aspects of the cell model
with binary Curie-Weiss interactions will be further discussed,
the main message of the present communication concerns the special
function $R(r;z)$ that normalizes the discrete Gauss-Poisson probability distribution function introduced in \eqref{GPD}.
We come up with several useful physical conclusions throughout the paper.
However, our primary goal is to bring
the essentially unknown mathematical objects \eqref{GPD} and \eqref{R} arising in a specific branch of theoretical physics to the broad mathematical community.
To the best of our knowledge, neither the discrete Gauss-Poisson probability distribution function $p_{GP}(n ;z,r)$ nor the special function $R(r;z)$ have been
considered in the mathematical literature before.

The function $R(r;z)$ seems to appear
for the first time in the physical paper \cite[p. 440]{PRT11e}
where the sum of exactly the same functional form as in \eqref{R} resulted from certain integration and started with $n=2$.
Later \cite[p. 20]{Rebenko13}, an analogous sum appears in even more general form, where the second power of $n$ in the summand is replaced by $n^m$.
In different disguises, the function $R(r;z)$ can be found
in a series of papers \cite[p. 810, (3.15)]{KDR15}, \cite[p. 5, (2.15)]{KKD20}, \cite[p. 3, (17)]{KD22}, \cite[p. 3, (9)]{DKRP24}, where the statistical mechanics calculations were based on grand canonical ensemble (see e.g. \cite[Sec. 2.4]{HansenMcDonald13}).
Apparently, the functional form of the function $R(r;z)$ in \eqref{R} should be characteristic to calculations using this kind of statistics; actually, its general shape is the same as that of the grand-canonical partition function $\Xi=\sum_{N=0}^{\infty}z^N\,Z_N/N!$ (see \cite[(2.4.6)]{HansenMcDonald13}
and \eqref{ZGR} below).
Apart from the sum $R(r;z)$ alone, for actual calculations of the mentioned papers
several lowest-order moments
\be\label{BEM}
M_m(z,r)=\left[R(r;z)\right]^{-1}\sum_{n=0}^\infty\;
n^m\;\frac{\e^{zn}}{n!}\,\e^{-\frac 12\,rn^2},\qquad\qquad m\in\mathbb N
\ee
of the distribution \eqref{GPD} have been needed.
However again, as far as we know, no mathematical properties of the sum $R(r;z)$
or the moments $M_m(z,r)$ have been studied in any physically-oriented work, apart
from their numerical evaluations using \emph{truncated} series of exponentials.

The truncation of the infinite series in \eqref{R} or \eqref{BEM} is indeed an important dangerous aspect, which could potentially lead to negative consequences.
The reason is that the large-$z$ behavior of a finite sum of exponential functions qualitatively differs (at any order of truncation) from the true
asymptotics of the infinite sum. We shall return to this issue in a subsequent publication.

\vspace{1mm}
Our motivation in the present work consists several aspects.

($i$) First of all, we are quite sure that it is of primary interest to initiate a mathematical investigation of the discrete Gauss-Poisson probability distribution function $p_{GP}(n ;z,r)$, its normalization $R(r;z)$, and, further, the
associated moments $M(r;z)$.
In doing so, we have determined the non-trivial asymptotic behavior of $R(r;z)$ along with some related functions and visualized an intriguing oscillating behavior of
$\ln R(r;z)$ in the vicinity of its  asymptotics. Of course, it would be a highly interesting challenge to derive any explicit determinations of these functions for some special sets of parameters, although such a goal seems to be rather hopeless for the moment.

($ii$) It is certainly expected that mathematical studies of this kind must have
appropriate physical consequences and implications.
An example is the issue of convergence of the master integral
\eqref{Y}, involving $\ln R(r;z)$ (see also \eqref{Z}),
in the marginal case of physical stability discussed in Sections \ref{SIS} and \ref{SII}. We hope that the list of such applications will be extended in the future.

($iii$) Our desire is to draw attention of mathematicians to the interesting and practically important mathematical objects discussed in the present communication and trigger their further investigation.

($iv$) On the other side, we observed a lack of explicit analytical results
in physical papers where the function $R(r;z)$ and moments $M_m(z,r)$ played an essential role, apart from a quite general proof \cite{KKD20} of the existence of phase transitions in underlying systems. All subsequent work, to be mentioned below, has been dedicated to numerical investigations. Here, we are starting to fill this gap, by presenting several instances of related analytical calculations and their implications.

\section{Physical context}
The special discrete Gauss-Poisson probability distribution function
\eqref{GPD} and associated infinite sums like \eqref{R} and \eqref{BEM}
appear in statistical mechanics investigations of simple classical fluids
within the framework of the grand canonical ensemble.
The physical basis for the present study is the cell model with binary Curie-Weiss interactions, introduced in \cite{KK16,KKD20} for the purpose of
modelling the simple fluids and describing phase transitions that occur in such
physical systems. Extended studies of this model have been subsequently done in
\cite{KKD18,KD22}, and, more recently in \cite{DKRP24,DKR24}.

In the next section, we are going to describe the underlying physical model while
closely following the reference \cite{KKD20}.

\subsection{Two-point interactions and their energy}\label{REX}

It is assumed that a classical fluid system, which may be in a gaseous or liquid state, occupies a macroscopic container of volume $V\subset\mathbb R^3$ in three space dimensions.
The volume $V$ is divided into $N$ non-overlapping congruent cubic cells $\Delta_\ell$ ($\ell=1,...,N$) of volume $v=V/N$. The system consists of a
\emph{variable} number $n\in\mathbb N$ of point-like particles with
three-dimensional coordinates $\{x_1,...,x_n\}$.
The particles can randomly occupy any cell $\Delta_\ell$ in the volume $V$.

The two-point interaction energy of particles with coordinates $x_i$ and $x_j$ is assumed to be
\be\label{CV}
\Phi_N(x_i,x_j)=-\,\frac{J_1}{N}+J_2\sum_{\ell=1}^N c_\ell(x_i)\,c_\ell(x_j)\,.
\ee
Here:
\vspace{-2mm}
\bi\itemsep1mm
\item[$\bm\cdot$] $J_1>0$ measures the strength of attraction between ANY two of $n$ particles in $V$
\item[$\bm\cdot$] $J_2>0$ is the strength of repulsion between the particles
\underline{inside of a cell $\Delta_\ell$}
\item[$\bm\cdot$] $c_\ell(x_i)$ are the occupation indicators for the cell $\Delta_\ell$;
they are defined via
\vspace{-2mm}
\bc
$c_\ell(x_i)=1$\quad if\quad $x_i\in\Delta_\ell$,\qquad
$c_\ell(x_i)=0$\quad if\quad $x_i\notin\Delta_\ell$\,.
\ec
\ei
\noi
Thus, the overall energy of the system of $n$ particles
with the Curie-Weiss interactions in the volume $V$ is given by
\be\label{OV}
W_N^{(n)}=\frac12\sum_{x_i,x_j}\Phi_N(x_i,x_j)=-\frac12\,\frac{J_1}N\,n^2+
\frac12\,J_2\sum_{x_i,x_j}\sum_{\ell=1}^N c_\ell(x_i)c_\ell(x_j)\,.
\ee

According to a rigorous result by Ruelle \cite{Ruelle70} (as quoted in \cite{KKD20}), the thermodynamic stability of the system under consideration is provided by the condition
\be\label{RUE}
\int_Vdx\,\Phi_N(x,y)>0\qquad\forall\,y\,\in V\,,
\ee
where the integration runs over the three-dimensional volume $V$ of the system, and we use a short-hand notation $dx=d^3x$.
With $\Phi_N(x,y)$ from \eqref{CV}, the condition \eqref{RUE} implies the requirement
$J_2/J_1>1$ \cite[(2.3) -- (2.4)]{KKD20}; see also \eqref{f}.
\label{ppr}

All thermodynamic properties of the system derive from
the grand-canonical partition function at the (absolute) temperature $T$,
\be\label{ZGR}
\Xi_N=\sum_{n=0}^\infty\,\frac{\zeta^n}{n!}\;Z_n\,,\qquad\qquad
Z_n=\int(dx)^n\exp\big[{-\beta W_N^{(n)}}\big]\,,
\ee
where the inverse temperature $\beta$ and activity (fugacity) $\zeta$ are
\be\label{LA3}
\beta=\frac1{k_B\,T}\MM{and} \zeta=\frac{\e^{\beta\mu_{\rm phys}}}{\Lambda^3}\,,
\ee
$k_B$ is the Boltzmann constant, $\mu_{\rm phys}$ is the physical chemical potential, and
$\Lambda$ is the de Broglie thermal wavelength.
From now on, we set $\Lambda:=1$ as in \cite{KKD20}.

\subsection{Integral representation of the grand-canonical partition function}
\label{RER}

An explicit calculation detailed in \cite{KKD20} yields a single-integral representation for the grand-canonical partition function $\Xi_N$ of the cell model with Curie-Weiss interactions. Using a parametrization slightly differing from that
accepted in \cite{KKD20}, we reproduce it in the form
\be\label{Y}
\Xi_N\equiv\Xi_N(p,r,\mu,v)=\sqrt{\frac N{2\pi p}}\int_{-\infty}^\infty dy\,
\e^{N E(p,r,\mu,v;y)}\,.
\ee
Here
\be\label{EE}
E(p,r,\mu,v;y)=-\frac{y^2}{2p}+\ln K(p,r,\mu,v;y)\,,
\ee
while
\be\label{KK}
K(p,r,\mu,v;y)=\sum_{n\ge0}\frac{v^n}{n!}\,\e^{(y+p\,\mu)n-\frac 12\,r n^2}=
R(r;y+p\,\mu+\ln v)
\ee
where we encounter just the function $R(r;z)$ from \eqref{R}, with a shifted argument.

\label{rule}
In the limit $N\to\infty$, we have
$$
\lim_{N\to\infty}N^{-1}\ln\Xi_N=E(p,r,\mu,v;\bar{y})\,,
$$
where
$E(\bar y)$ is the maximum value of $E(y)$ at the maximum point $\bar y$, which is defined as a solution to the extremum condition $E'(y)=0$, provided that
$E''(\bar y)<0$.
The limit $N\to\infty$ employed in \eqref{Y} for evaluation of this integral via the Laplace method is directly related to the thermodynamic limit, which is performed
at a fixed cell volume $v$. This drives the system's volume $V$ to infinity, while
$v$ can be fixed at $v=1$ in subsequent calculations.

At this point, it is important to stress the similarities and differences in variables and parameters employed in \cite{KKD20} and throughout the current paper.

\label{dif}
The two independent fundamental thermodynamic variables in the problem are the inverse temperature
$\beta\ge0$ and the physical chemical potential $\mu_{\rm phys}\in\mathbb R$.
As before, the dimensionless normalized inverse temperature is defined as
$\underline{p=\beta J_1}\ge0$.
Its counterpart involving the interaction parameter $J_2$
is $\underline{r=\beta J_2}\ge0$.
These variables are proportional to one another, and their ratio is
\be\label{f}
\frac r p=J_2/J_1\equiv a>1 \mm{in notation of  \cite{KKD20},\quad and}
\frac r p=J_2/J_1\equiv f>1 \mm{in notation of \cite{KD22}.}
\ee
Basically, \emph{any one} of the quantities $p$ and $r$ can be chosen to represent the normalized inverse temperature.
In \cite{KKD20}, the role of such a variable has been assigned to $p$.
Consequently, the remaining temperature-like quantity $r$ had to be represented as $r=p\,\frac r p=p\,a$. The inequality $a=f>1$ is a direct consequence of the thermodynamic stability condition \eqref{RUE}.

Moreover, the dimensionless combination $\beta\mu_{\rm phys}$ appearing in the fugacity \eqref{LA3} is a product of two independent physical variables.
If the quantity $p$ is chosen to represent the temperature as in \cite{KKD20}, we write
$\beta\mu_{\rm phys}=(\beta J_1){\cdot}\,\mu_{\rm phys}/J_1\equiv p\,\mu$,
where we define the dimensionless normalized chemical potential $\mu\in\mathbb R$ via \underline{$\mu\equiv\mu_{\rm phys}/J_1$}.

\vspace{1mm}
We emphasize that our parametrization of physical variables related to the temperature
($p=\beta J_1$) and chemical potential ($\mu=\mu_{\rm phys}/J_1$) differs from that
accepted in \cite{KKD20}.
In fact, in this reference the inverse temperature $\beta$ appears in the "basic set of thermodynamic variables" $(p,\mu)$ two times, both in $p$ and in $\mu$,
due to the definition of $\mu=\beta\mu_{\rm phys}$ (see \cite[p. 3]{KKD20}), which combines two physically distinct and independent variables
$\beta$ and $\mu_{\rm phys}$ into one.

\section{Some simple special cases}\label{SSS}

It seems that in the previous publications on the cell model
described in Section \ref{REX}
no simple limiting cases have been considered, such as the high-temperature limit $\beta\to0$ with finite interactions $J_1,\,J_2>0$, or the ideal gas with vanishing interactions $J_1=J_2=0$ at arbitrary temperature $\beta$.
These special cases would require calculations of the grand partition function
$\Xi_N(p,r,\mu,v)$ in the limit of the vanishing parameter $p=\beta J_1$.
However, it is not evident how to proceed with this limit starting from the integral representation \eqref{Y} with the function $E$ given in \eqref{EE} in the integrand.

To avoid the apparent difficulty with the $p\to0$ limit in \eqref{Y}, we slightly modify this integral by scaling the integration variable via $y=t\sqrt p$\,.
Moreover, we find it helpful to use the explicit expressions for the normalized quantities $p$, $r$ and $\mu$ introduced in the previous section.
Thus, we write the integral representation \eqref{Y} in the form
\be\label{T}
\Xi_N(J_1,J_2;\beta,\mu_{\rm phys};v)=\sqrt{\frac N{2\pi}}\int_{-\infty}^\infty\!dt
\exp\Big\{N \Big[-\frac{t^2}2+\ln\sum_{n\ge0}\frac{v^n}{n!}\,
\e^{(t\sqrt{\beta J\,}_{\!1}\,+\beta\mu_{\rm phys})n-\frac 12\,\beta J_2n^2}\Big]\Big\}.
\ee

Hence, in the case of a noninteracting ideal gas with $J_1=J_2=0$ and arbitrary
finite $\beta<\infty$ and $\mu_{\rm phys}<\infty$, we remain with a simple
summation in \eqref{T} and obtain the
grand-canonical partition function $\Xi_N(0,0;\beta,\mu_{\rm phys};v)$ as
\be\label{TRE}
\Xi_N^{(id)}(\beta,\mu_{\rm phys},v)=\sqrt{\frac N{2\pi}}\int_{-\infty}^\infty\!dt
\exp\Big\{N\Big[-\frac{t^2}2+\ln\sum_{n\ge0}\frac{v^n}{n!}\,
\e^{\beta\mu_{\rm phys} n}\Big]\Big\}
=\exp\left(Nv\,\e^{\beta\mu_{\rm phys}}\right).
\ee
Thus, $\ln\Xi_N^{(id)}(\beta,\mu_{\rm phys},v)=V\e^{\beta\mu_{\rm phys}}$ in agreement (up to the factor $\Lambda^{-3}$, see \eqref{LA3})
with the well-known classical result, which can be found, for example, in
\cite[p. 395, (A2.11)]{Hill56}.

\vspace{2mm}
The high-temperature limit $\beta=0$ becomes trivial:
$$
\ln\Xi_N(J_1,J_2;0,\mu_{\rm phys};v)=\ln\Xi_N^{(id)}(0,\mu_{\rm phys},v)
= V\left.\e^{\beta\mu_{\rm phys}}\right|_{\beta=0}=V.
$$

\vspace{1mm}
The special case $J_1=J_2=0$ is interesting because of of the thermodynamic stability
requirement \eqref{RUE} as the latter implies the inequality $f=J_2/J_1>1$
(see \eqref{RUE}, \eqref{f}).
When $J_1=J_2=0$, we encounter the extreme limit of the equality
case $J_1=J_2$ corresponding to the stability edge given by $f=1$.
The existence of this limit is not anticipated by the condition \eqref{RUE},
but nevertheless, we just obtained a sensible and expected result in this marginal case.
This observation suggests that it would be interesting to consider the special
case of equal non-vanishing attraction and repulsion interactions $J_1=J_2>0$ in \eqref{CV}, that is $f=1$ and $0<r=p<\infty$ (see Sec. \ref{RER} around the equation \eqref{f}).

Another limiting case of strong repulsion interactions within the cells, $J_2\to\infty$, with an arbitrary and independent attraction parameter $J_1>0$ will be considered separately in Sec. \ref{J2B}.

And now, we will consider the convergence of equivalent basic integral representations \eqref{Y} or \eqref{T} on the edge of the thermodynamic stability
$J_1=J_2$.

\section{The issue of convergence of the integral \eqref{T} at the edge of stability}\label{SIS}

Having solved the problem related to the special case $p=0$ in the preceding section,
we can safely return to the integral representation \eqref{Y} to consider finite parameters $0<p<\infty$ .
Thus, following \cite{KD22}, we shift the integration variable via $y=z-p\mu-\ln v$.
Our starting point becomes the integral
\be\label{Z}
\Xi_N(p,r,\mu,v)=\sqrt{\frac N{2\pi p}}\int_{-\infty}^\infty\!dz
\exp\Big\{N \Big[-\frac1{2p}(z-p\,\mu-\ln v)^2+\ln R(r;z)\Big]\Big\},
\ee
where the function $R(z;r)$ appears as it is announced in \eqref{R}.
In comparison to \eqref{Y}, in the present representation
\eqref{Z}, the sum $R(r;z)$ under the logarithm contains \emph{only one}
physical parameter, namely the temperature-like thermodynamic variable $r=\beta J_2$.
All remaining "physics" enters the first simple quadratic term.

Completing the square in the exponents of summands in $R(r;z)$, we obtain
\be\label{CS}
zn-\frac 12\,rn^2=\frac{z^2}{2r}-\frac r2\,\Big(n-\frac zr\Big)^2.
\ee
Here, the first term on the right-hand side is the maximal value of the quadratic form on the left, for any $n\ge0$. Hence, for the grand-canonical partition function $\Xi_N$ in \eqref{Z} we can write
\be\label{YR}
\Xi_N\propto\e^{-\frac N{2p}\,(p\,\mu+\ln v)^2}\!\int_{-\infty}^\infty dz
\exp\Big\{N\Big[-\frac{z^2}{2p}\,\left(1-b\right)
+(\mu+p^{-1}\ln v)\,z+\ln \hat R(r;z)\Big]\Big\},
\ee
where we defined the ratio $\displaystyle{b:=\frac pr=\frac{J_1}{J_2}}$
in the range $b\in[0,1]$ (cf. \eqref{f}) and the modified $R$-sum (cf. \eqref{R})
\be\label{HR}
\hat R(r;z):=\sum_{n\ge0}\,\frac1{n!}\,\e^{-\frac r2\,\left(n-\frac zr\right)^2}\,.
\ee

We can observe that through the procedure of completing the square in \eqref{CS},
the value $\displaystyle{\e^\frac{z^2}{2r}}$ of the greatest summand in \eqref{R} appears, defining the leading term of the asymptotic behavior of the sum $R(r;z)$ for large $z$ (cf. \eqref{ARS}).
To determine further, sub-leading terms of the asymptotic expansion of $R(r;z)$,
one would have to employ more elaborated techniques described in
\cite{deBruijn,BO78,FS,Paris11}.

For the moment, we just notice that while the function $\ln R(r;z)$ roughly behaves as
\be\label{o}
\ln R(r;z)\sim\frac{z^2}{2r}+o(z^2) \MM{when} z\to\infty,
\ee
this implies that $\ln\hat R(r;z)\sim o(z^2)$ in the same limit.
For the integral \eqref{YR} this means that its convergence at large $z$
is controlled by the first quadratic term $\propto-z^2(1-b)$ for all $b<1$.
In particular,
\vspace{-3mm}
\bi\itemsep-1mm
\item[$\bm\cdot$]
the inequality $b<1$ ensures the convergence of the integral over $z$ in \eqref{YR},
in agreement with \eqref{RUE} and \eqref{f};
\item[$\bm\cdot$]  with $b>1$, the integral would diverge, which corresponds to an unphysical situation;
\item[$\bm\cdot$] the marginal case $b=1$ requires a special consideration taking into account the asymptotic behavior of the function $\ln\hat R(r;z)$ at large $z$.
\ei

In fact, in the absence of the $\propto z^2$ term in \eqref{YR}, the
convergence of the integral at large $z$ is controlled by the leading
term of the asymptotics of $\ln\hat R(r;z)$, that is, by the sub-leading $o(z^2)$
term in the asymptotic expansion of $\ln R(r;z)$ implied by \eqref{o}.

In the following sections, we will present the asymptotic estimates $R^{(\rm as)}(r;z\gg1)$ for the function $R(r;z)$ at large arguments $z$.
This will allow us to answer the question about the convergence
of the integral representation \eqref{Z} for the grand-canonical partition function
$\Xi_N(p,r,\mu,v)$ in the controversial marginal case $r=p$.

\section{The function $R(r;z)$ and its asymptotics}

The function
\be\label{RR}
R(r;z)=\sum_{n\ge0}\frac{\e^{zn}}{n!}\,\e^{-\frac 12\,rn^2}
\ee
is well defined for any $r\ge0$ and $z\in\mathbb R$.

At $r=0$, the exact result of the summation in \eqref{RR} is $R(0;z)=\exp(\e^z)$,
while in the limit $r\to\infty$ only the first term of the sum survives and thus we have $R(\infty;z)=1$.
Hence, for any $0\le r<\infty$, the function $R(r;z)$ lies within the bounds
$1\le R(r;z)<\exp(\e^z)$ and therefore $0<\ln R(r;z)<\e^z$.
Moreover, for $r>0$, the estimate from above
$R(r;z)\le\e^{z^2/(2r)+1}$ has been used in \cite{KKD20}.

In the following, we shall be interested in the asymptotic
behavior of the function $R(r;z)$ for finite $0<r<\infty$ and $z\gg 1$.

\subsection{The asymptotic behavior of $R(r;z)$}\label{RRR}

In order to obtain an asymptotic evaluation of the sum \eqref{RR}
in the limit $z\to\infty$ we employed a discrete analogue of the Laplace method
(see e.g. \cite{deBruijn,Copson,BO78,Fedoryuk89,Wong,FS,Paris11book,Temme})
commonly used for asymptotic approximations of integrals.

By contrast to the case of the asymptotic analysis of integrals, the literature on
its discrete version for sums is very scarce.
The discrete Laplace method is telegraphically outlined in the well-known books by
de Bruijn \cite[Ch. 3]{deBruijn}, Bender and Orszag \cite[p. 304 -- 305]{BO78}, and
Flajolet and Sedgewick \cite[p. 761 -- 762]{FS}.
A detailed development of the discrete Laplace method capable of producing the
next-to-leading terms of the asymptotic expansion has been performed in the paper
by Paris \cite{Paris11}, where several additional relevant references can also be found. However, its exposition is based on a rather specific example and seemingly does not provide any explicit receipt for treating the asymptotic behavior of generic discrete sums.

For the sum \eqref{RR} we have found the asymptotic formula
\be\label{ASW}
R^{(\rm as)}(r;z\gg1)\sim\frac{\e^{-\frac r2\,\m^2}}{\sqrt{1+r\m}}\,
\left(\frac{\e^{z+1}}\m\right)^\m,
\quad\MM{where}\m=\frac1r\,W\left(r\e^z\right)
\ee
and $W(x)$ is the Lambert $W$-function, the solution to the equation $W\e^W=x$.
An exiting introduction to the Lambert $W$-function can be found in the excellent paper
\cite{Lambert}.

The asymptotic formula \eqref{ASW} holds for arbitrary finite $r\in\mathbb R_+$,
\emph{including $r=0$}, in which case it reproduces the exact result
$R(0;z)=\exp(\e^z)$ valid for any $z<\infty$. This can be easily checked by using the Taylor expansion  of the Lambert $W$-function $W(x\to0)=x+O(x^2)$ \cite[(3.1)]{Lambert}.

Shifting the summation index in $R(r;z)$ and using \eqref{ASW} we derived the asymptotic estimate for the function
$\displaystyle{R_1(r;z)\equiv \frac d{dz} R(r;z)}$:
\be\label{ASW1}
R_1^{(\rm as)}(r;z\gg1)\sim\frac{\e^{-\frac r2\,\m_1^2}}{\sqrt{1+r\m_1}}\,
\left(\frac{\e^{z+1-r}}{\m_1}\right)^{\m_1}\!\!\cdot\,\e^{z-r/2}
\quad\MM{with}
\qquad \m_1=\frac1r\,W\left(r\e^{z-r}\right).
\ee
Again, the exact result $R_1(0;z)=\exp(\e^z)\,\e^z$ is reproduced from the last equation.

The asymptotic formula \eqref{ASW} can be simplified to the form involving only elementary functions. The result is, with the notation\; $\zeta\equiv z/r$,
\be\label{ARS}
r\ln R^{(\rm as)}(r>0;z\gg1)\sim
\frac12\,z^2-z(\ln\zeta-1)+\frac12\,\ln^2\zeta-\frac r2\,\ln z+
O\Big(\frac{\ln\zeta}z\Big)\,.
\ee
As we see, the asymptotic expansion \eqref{ARS} includes all divergent terms
along with a constant, while all descending terms are discarded.
It is also observed that the leading term of the asymptotic expansion of
the function $r\ln R(r;z)$ does not  depend on $r$.

\vspace{2mm}
Furthermore, the evaluation \eqref{ARS} implies, for finite $0<r<\infty$,
\be\label{ARS1}
r\,\frac{d\ln R(r;z)}{dz}=r\,\frac{R_1(r;z)}{R(r;z)}\sim
z-\ln\zeta+\frac1{2z}\,(2\ln\zeta-r)+O\Big(\frac{\ln\zeta}{z^2}\Big),\quad
z\gg1\,.
\ee
The information of this kind can be interesting in the context of studying the
moments \eqref{BEM} of the discrete Gauss-Poisson probability distribution function
\eqref{GPD}.

In deriving the simplified asymptotics \eqref{ARS}, we employed the asymptotic expansion of the Lambert function $W(\e^z)$ at $z\to\infty$
(cf. \cite[(4.19)]{Lambert}):
\be\label{WAS}
W(\e^z)\sim z-\ln z+\frac{\ln z}z+O(z^{-2}\ln^2z)\,.
\ee
Since we used this expansion for the function $\m r=W(r\e^z)$ from \eqref{ASW} at large argument $r\e^z$, the alternative limit $r\to0$ has been lost in the results \eqref{ARS} and \eqref{ARS1}.

\subsection{Graphical illustrations}

In the present section, we show a few plots related to the function $R(r;z)$ and its
asymptotic behavior.

\vspace{4mm}
\hspace{-7mm}\includegraphics[width=8.5cm,height=5.6cm]{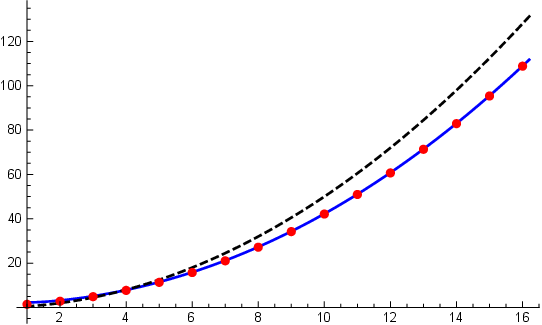}\quad\!
\includegraphics[width=8.4cm,height=5.6cm]{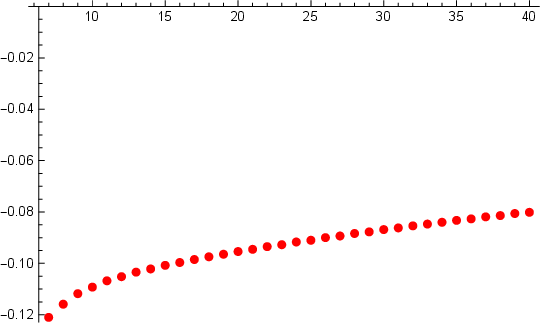}
\vspace{2mm}

\noi
Fig 1. LEFT: Red dots: the function $r\ln R(r;z)$ at $r=1.8$ and $z=1,...,16$.
The solid blue curve represents the asymptotics $r\ln R^{(\rm as)}(r;z\gg1)$ of the same function
given by the formula \eqref{ARS}. The dashed black curve shows the leading asymptotic term $z^2/2$ of $r\ln R^{(\rm as)}(r;z\gg1)$ from \eqref{ARS}.
RIGHT: The difference \eqref{DIF} between the function $r\ln R(r;z)$ and its asymptotics from \eqref{ARS} at $r=1.8$.

\vspace{2mm}
Figure 1 represents a comparison of the numerical calculation of the function
$r\ln R(r;z)$ (performed with the help of Mathematica \cite{Math12}) and the analytical formula \eqref{ARS} for its asymptotics at $r=1.8$.
As we see, the asymptotic evaluation \eqref{ARS} compares very well with the full function $r\ln R(r;z)$.
Moreover, as it often happens in the asymptotic analysis,
the asymptotic formula derived under condition $z\to\infty$ gives a good approximation
for the underlying function even for rather small arguments $z$, like $z=2$ or $3$, which
do not actually meet this requirement. The leading asymptotic term $z^2/2$, accessible in an
elementary way by completing the square in exponents of summands in $R(r;z)$
(see \eqref{CS}) does not provide such a good fit as the formula \eqref{ARS}.

In the above example we have taken a rather small value of the parameter $r$.
In such a case, deviations of the original function $r\ln R(r;z)$ from its asymptotics are very small, and their difference
\be\label{DIF}
r\ln R(r;z)-r\ln R^{(\rm as)}(r;z\gg1)
\ee
shows a smooth monotonic behavior.

The situation changes drastically when we move towards larger values of the parameter $r$ and exceed certain threshold value $r=r^*$.
If we consider such $r>r^*$, the deviations between the function
$r\ln R(r;z)$ and its asymptotics become much more essential, and their difference
\eqref{DIF} becomes oscillatory. Such situation is illustrated by the couple of graphs in Figure 2.

\vspace{4mm}
\hspace{-5mm}\includegraphics[width=8cm,height=5cm]{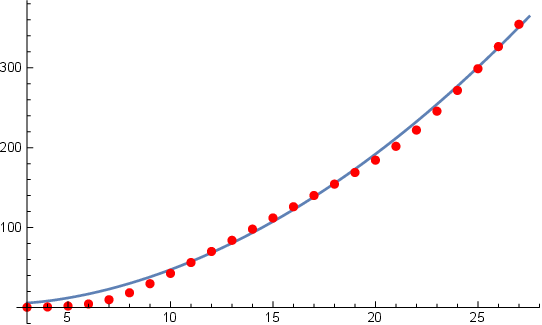}\quad
\includegraphics[width=8cm,height=5cm]{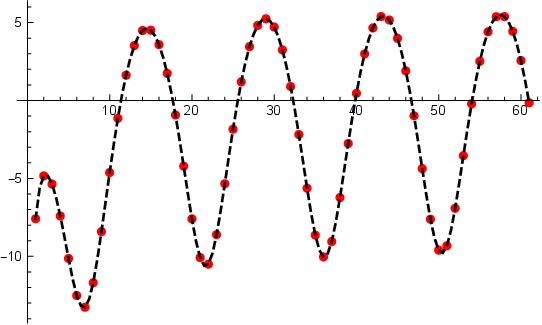}
\vspace{4mm}

\noi
Fig 2. LEFT: Red dots: the function $r\ln R(r;z)$ at $r=14$ and $z=3,...,27$.
The solid blue curve represents the asymptotics $r\ln R^{(\rm as)}(r;z\gg1)$ of this function
given by the formula \eqref{ARS}.
RIGHT: Red dots: The difference \eqref{DIF} between the function $r\ln R(r;z)$ and its asymptotics from \eqref{ARS}. The dashed curve provides an eye-guide.
The argument $z$ is extended here up to $z=61$.

\vspace{2mm}
The functions $R(r;z)$ and $r\ln R(r;z)$, and their non-trivial behavior in the vicinity of asymptotics certainly deserve further investigation.
Nevertheless, at the moment we return to the convergence issue raised in Sec. \ref{SIS}.

\section{Back to the integral \eqref{Z} at the stability edge}\label{SII}

In Sec. \ref{SIS}, we have seen that the convergence of the integral \eqref{Z},
and thus, the stability of the underlying physical system, strongly depend on the
value of the parameter  $\displaystyle{b:=\frac pr=\frac{J_1}{J_2}}$ introduced in
\eqref{YR}. While it was evident that in \eqref{YR}, the integral over $z$ converges
if $b<1$ and diverges if $b>1$, the question of its behavior in the marginal case
$b=1$ remained open. Now, possessing the required information on the large-$z$ behavior of its integrand, we are in a position to give a solution to this problem.

Let us write the integral over $z$ in \eqref{YR} as
\be\label{BI}
\int_{-\infty}^\infty\!dz\,\e^{N\,E(p,r;z)}\MM{with}
E(p,r;z)=-\frac{z^2}{2p}\,\left(1-b\right)+(\mu+p^{-1}\ln v)\,z+\ln\hat R(r;z)
\ee
and the function $\hat R(r;z)$ defined in \eqref{HR}.
At $b=1$, the quadratic term in $E(p,r;z)$ disappears, and we set $r=p$.
Thus we remain with the integrand
\be\label{E0}
E(p,p;z)=(\mu+p^{-1}\ln v)\,z+\ln\hat R(p;z)\,.
\ee

By the definition of the function $\hat R(r;z)$ (see \eqref{CS}, \eqref{HR}),
the asymptotic behavior of $\ln\hat R(p;z)$ at large argument $z$ is given by
\be
\ln\hat R^{(\rm as)}(p;z\gg1)=\ln R^{(\rm as)}(p;z\gg1)-\frac{z^2}{2p}\sim
-\frac zp\,(\ln z-\ln p-1)+O(\ln^2z)\,,
\ee
where the expression on the right-hand side results from the truncated asymptotic
estimate \eqref{ARS} with the replacement $r\mapsto p$. Thus, for large enough $z$ we have
\be\label{E01}
E(p,p;z\gg1)\sim-\,\frac zp\,\ln z+
\,\frac zp\left(p\,\mu+\ln v+\ln p+1\right)+O(\ln^2z)\,.
\ee
We see that the negative next-to leading asymptotic term $\sim-z\ln z$ of the function
$p\ln R^{(\rm as)}(p;z\gg1)$ from \eqref{ARS} guarantees the convergence of the integral \eqref{BI} at $z\to+\infty$.

This means that in the special case $J_1=J_2$ of \eqref{CV}, the mathematical problem to solve does not encounter any convergence problem, and thus can lead to meaningful results from a physical point of view. Its numerical solution may go along the lines of \cite{KKD20,KKD18}, and \cite{KD22}, where only strictly smaller than unity values of $b$ have been taken into account: See, for example, \cite[Table 1]{KKD20},
\cite[p. 249]{KKD18} where a close-to-the-edge value $a=b^{-1}=1.0001$ has been considered among the others.

\subsection{Explicit solution of the $b=1$ problem in the asymptotic limit}

Now, we are in a position to evaluate the integral \eqref{BI} in the asymptotic limit, that is with the function $E(p,p;z\gg1)$ from \eqref{E01} in the integrand. Let us write this function in the form
\vspace{-3mm}
\begin{flalign}\nonumber&
E(p,p;z{\gg}1)=p^{-1}\left(\hat\mu z-z\ln z\right)\mm{with the short-hand notation}
\hat\mu\equiv p\,\mu+\ln v+\ln p+1\,.
&\end{flalign}
\vspace{-8mm}

As described after the equation \eqref{KK}, we have the extremum condition
$$
E_0'(p,p;z\gg1)=p^{-1}\left(\hat\mu-1-\ln z\right)=0.
$$
For finite $p>0$, hence follows\;
$\ln\bar z=\hat\mu-1$, and $\bar z=\e^{\,\hat\mu-1}$\; for the extremum position.
The second derivative $E_0''(p,p;z\gg1)=-1/(p\,z)<0$ as it should be within the Laplace method, indicating that the argument $z=\bar z$ indeed corresponds to a maximum.
The maximum's height is then
$$
E(p,p;\bar z\gg1)=p^{-1}\left[\hat\mu\e^{\,\hat\mu-1}-\e^{\,\hat\mu-1}(\hat\mu-1)\right]=
p^{-1}\e^{\,\hat\mu-1}=p^{-1}\e^{p\,\mu+\ln v+\ln p}=v\,\e^{p\,\mu}
=v\,\e^{\beta\mu_{\rm phys}},
$$
which evaluates the integral in \eqref{BI} and completely agrees with the ideal-gas result derived in the special case $r=p=0$ before, see \eqref{TRE}.

It would be interesting to see whether the inclusion of more terms of the asymptotic expansion for the function $E(p,p;z\gg1)$ or employing the numerical analysis along the lines of \cite{KD22} will produce deviations from the ideal-gas result obtained for the marginal case $b=1$ just above.

\section{The strong-repulsion limit $J_2\gg J_1$}\label{J2B}

In this section, we shall consider the limiting case of strong repulsion interactions within the cells, that is $J_2\to\infty$, with arbitrary finite strength of attraction between particles $J_1>0$ (see Sec. \ref{REX}).

Our starting point will be the integral representation \eqref{Y}-\eqref{KK} where the large repulsion
parameter $J_2$ implicitly appears through the quantity $r=\beta J_2$.
Following \cite{KKD20}, we shall control the temperature using the dimensionless
variable $p$, and for the parameter $r$ we write $r=(\beta J_1)J_2/J_1=p\,f$
(cf. \eqref{f}).
For our present purposes, the relation $J_2\gg J_1$ will be controlled by the
ratio $f\equiv J_2/J_1\gg1$. With this in mind, we approximate the function
$K(p,r,\mu,v;y)$ in \eqref{KK} via
\be\label{FUS}
K(p,r,\mu,v;y)\simeq\sum_{n=0}^1\frac{v^n}{n!}\,\e^{(y+p\,\mu)n-\frac 12\,r n^2}=
1+v\,\e^{y+p\,\mu-\frac12\,pf}+O(\e^{-2pf})\,,
\ee
that is, we take into account only the first non-trivial term in the sum \eqref{KK}.
Without this exponential contribution, we would obtain a quite trivial result for
the limit $f=\infty$.

Thus, we are going to consider the
grand-canonical partition function $\Xi_N(p,f\gg1,\mu,v:=1)$ given by the integral
(cf. \eqref{Y}-\eqref{KK})
\be\label{BIF}
\int_{-\infty}^\infty\!dy\,\e^{N\,E(y)}\MM{with}
E(y)=\ln\left(1+\e^{y+p\,\mu-\frac12\,pf}\right)-\frac{y^2}{2p}\,.
\ee

\subsection{A simple way to the critical point}\label{SWY}

To simplify the following expressions we introduce short-hand notations
\be\label{SHN}
y+p\,\mu-\frac12\,pf\equiv u(y)\MM{and}\e^{u(y)}\equiv x>0\,.
\ee
It is conceivable that there is a special point
$y=y_0=p(\frac12f-\mu)$, for which
$u(y_0)=0$ and $\e^{u(y_0)}\equiv x_0=1$.

\vspace{2mm}
To begin with, let us calculate the first derivative of the function $E(y)$ from \eqref{BIF}:
\be\label{POE}
E'(y)=\frac{\e^{u(y)}}{1+\e^{u(y)}}-\frac yp=\frac{p\,x-y-yx}{p(1+x)}\,.
\ee
Note that by the definition of the function $x$ in \eqref{SHN}, we have
$\displaystyle{\frac{dx}{dy}=\frac{d\,\e^{u(y)}}{dy}=\e^{u(y)}=x}$.

At the special point $y=y_0$, with $x_0=1$ in \eqref{POE}, we have
\be\label{POC}
E'(y_0)=\frac{p-2\,y_0}{2p}\,,
\ee
and $E'(y_0)$ vanishes when $y_0=\bar y_0=p/2$. By the definition of $y_0$ as
$\displaystyle{y_0=\frac p2\,(f-2\mu)}$, the extremum condition $\bar y_0=p/2$ implies the
relation
\be\label{MUC}
\mu=\mu_c=\frac12(f-1)
\ee
among the underlying physical parameters of the problem.

According to the rules for calculating the integral over $y$ in \eqref{BIF} using the
Laplace method for large $N$ given in page \pageref{rule}, we have to check the
sign of the second derivative $E''(y)$ at the extremum point $y=\bar y_0$.
This second derivative has to be negative to provide that the function $E(y)$ has a
\emph{maximum} at $y=\bar y_0$.

A straightforward calculation yields for the second derivative of $E(y)$
\be\label{POD}
E''(y)=\frac{\e^{u(y)}}{(1+\e^{u(y)})^2}-\frac1p=
-\,\frac{x^2+(2-p)\,x+1}{p(1+x)^2}\,.
\ee
In our special situation with $y=y_0$, we have
\be\label{POF}
E''(y_0)=\frac{p-4}{4p}\,.
\ee
This means that $E''(y_0)<0$ for any $y_0$ if $p<p_c=4$, including the extremum point
found at $y_0=\bar y_0=p/2$, which thus appears to be a simple maximum. And, at the critical value of $p$,  $p=p_c=4$, the
second derivative $E''(y_0)$ vanishes, thus indicating some more complicated behavior
of the function $E(y)$.

To see the fate of the maximum $E(\bar y_0)$ at $p=p_c$, we have to calculate the third and the fourth derivatives of $E(y)$. These are
\be\label{POT}
E'''(y)=\frac{x(1-x)}{(1+x)^3}\MM{and}E^{iv}(y)=x\,\frac{(1-x)^2-2\,x}{(1+x)^4}\,,
\ee
and it is noticable that these higher derivatives have no explicit dependence on $p$
and $y$ in contrast to $E'(y)$ and $E''(y)$ in \eqref{POE} and \eqref{POD}.
With $x=x_0=1$,
\be\label{POW}
E'''(y_0)=0\MM{and}E^{iv}(y)=-\,\frac18\,.
\ee
This means that at $p=p_c=4$, the simple maximum at $y=\bar y_0=p/2$ and $p<4$ transforms into a flat degenerate maximum of the function
\be\label{ECY}
E_c(y)=\ln\left(1+\e^{y-2}\right)-\frac{y^2}8
\ee
at $y=y_c=2$.
The maximum value of this function is $E_c(y_c)=E_c(2)=\ln2-1/2$, which is a special
case of \eqref{MIP} at $p=4$.

Apart from the basic extremum condition $E'(y)=0$, the conditions $E''(y)=0$ and $E'''(y)=0$ required at the critical point, are simplified analogs of equations
\cite[(28)]{KD22} used in this reference for a numerical determination of critical points of the underlying physical system.

We also observe that the equation $E'''(y)=0$ with $E'''(y)$ given in \eqref{POT} immediately yields the non-trivial solution $x_0=1$, and thus $u(y_0)=0$ chosen by
inspection at the very beginning of the present section.

In our present setting with $f\gg1$, we have found a critical point occurring at
the (inverse) critical temperature $p=p_c=4$ and the critical value of the chemical potential $\mu=\mu_c=\frac12(f-1)$.
The value $p_c=4$ agrees with that found numerically in \cite[p. 13, Table 1]{KKD20}
for $f=a=10$. With this value of the parameter $f$, our calculation leads to the
value $\mu_c(10)=9/2$, while there is no information on its counterpart in \cite{KKD20}. The critical value $p_c=4$ agrees also with that found in
\cite[p. 249]{KKD18} for $a=10$, while some discrepancy exists in evaluations of $\mu_c$. Our value $\mu_c(10)=9/2$ must be multiplied by $p_c=4$ to fit the notation
$\mu_c$ accepted in \cite{KKD18} (see p. \pageref{dif}). This would lead to
$\mu_c^{10}=18$ in notation of \cite{KKD18}, while the numerical result of this reference is $\mu_c^{10}\simeq15.5$, the value of the same order of magnitude, but not quite the same; the difference being about $14\%$.

\subsection{Temperature dependence of $E(y)$ at $\mu=\mu_c$}\label{SPE}

At the critical value of the chemical potential, $\mu=\mu_c=(f-1)/2$ (see \eqref{MUC}), the function $E(y)$ from \eqref{BIF} reduces to
\be\label{BIP}
E(\mu_c;y)=\ln\left(1+\e^{y-p/2}\right)-\frac{y^2}{2p}\,.
\ee
The dependence on the large parameter $f$ has disappeared,
and the sole remaining physical parameter in \eqref{BIP} is the inverse temperature $p$. Different shapes of the function $E(\mu_c;y)$ at different values of the parameter $p$, namely $p<p_c$, $p=p_c$, and $p>p_c$, are shown in Fig. 3:

\vspace{4mm}
\begin{minipage}{8cm}
\includegraphics[width=8cm,height=6cm]{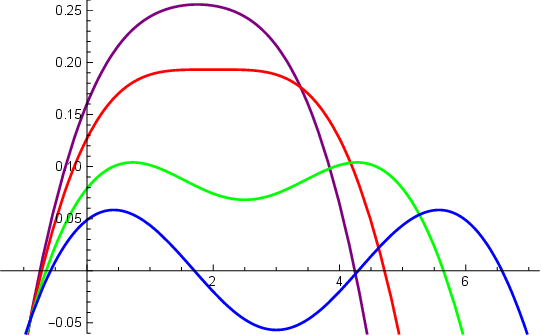}
\end{minipage}
\hfill
\begin{minipage}{8cm}
Fig. 3. Plots of the function $E(\mu_c;y)$ from \eqref{BIP} at four different values of the parameter $p$\,: $p=3.5<p_c$ (purple), $p=4=p_c$ (red), $p=5>p_c$ (green), and $p=6$ (blue).
One can see that the purple maximum occurs at $\bar y_3=1.75$, the red one ---  at $\bar y_4=2$, and the green and blue minima --- at $\bar y_5=2.5$ and $\bar y_6=3$, respectively.
Moreover, all curves are symmetric with respect to their extremum positions $\bar y_p$.
The blue curve with $p=6$ is a counterpart of the blue one in Figure 4 of \cite[p. 14]{KKD20}.
\end{minipage}
\vspace{4mm}

The plot of $E(\mu_c;y)$ at $p=6$ is analogous to that given in \cite[p. 14, Fig. 4]{KKD20},
and the both are very similar even in numerical values, though the latter one is produced
with the infinite sum over $n$ in \eqref{FUS}, and a different value of $\mu_c$.

Vanishing of the first derivative of $E(\mu_c;y)$ given by \eqref{POE} with
$u(y)=y-p/2$ yields the extremum condition
\be\label{EXC}
\e^{y-p/2}=\frac y{p-y}\,.
\ee
An evident trivial solution to the last equation is $\bar y_0=p/2$, the same as that
following from \eqref{POC}. As far as $p<p_c=4$, the coordinates $\bar y_0=p/2$
are the positions of simple maxima of $E(\mu_c;y)$ represented by the purple curve in Fig. 3. At $p=p_c=4$, the maximum becomes degenerate and acquires the flat form presented by the red curve in this figure.
Further, when we move to $p>p_c>4$, the extremum at $\bar y_0=p/2$ becomes a
\emph{minimum}.
Two side maxima of equal height appear at the same time, and this situation is
illustrated by the green and blue curves.
The positions of these maxima are defined by two further solutions $\bar y_1$
and $\bar y_2$ to the extremum equation \eqref{EXC}, apart of $\bar y_0=p/2$.

To confirm the apparent symmetry of the curves in Fig. 3, we shift in $E(\mu_c;y)$ the variable $y$
via $y-p/2=s$. Thus we obtain the function
\be\label{BIS}
\hat E(\mu_c;s)=\ln\left(1+\e^s\right)-\frac1{2p}\,\Big(s+\frac p2\Big)^2,
\ee
which becomes symmetric with respect to the ordinate.
To see that $\hat E(\mu_c;s)$ is indeed an even function, let us change $s\mapsto-s$
in \eqref{BIS}. Thus, we obtain
\begin{align}\nn
&
\hat E(\mu_c;-s)=\ln\left(1+\e^{-s}\right)-\frac1{2p}\,\Big(-s+\frac p2\Big)^2=
\ln\left[\e^{-s}(1+\e^s)\right]-\frac1{2p}\,\Big(s-\frac p2\Big)^2=
\\\nn&
=\ln\left(1+\e^s\right)-\frac1{2p}\Big[\Big(s-\frac p2\Big)^2+2ps\Big]
=\ln\left(1+\e^s\right)-\frac1{2p}\,\Big(s+\frac p2\Big)^2=\hat E(\mu_c;s)\,.
\end{align}

In Fig. 4 we plot the function $\hat E(\mu_c;s)$ in the same four cases as in Fig. 3. The whole picture becomes symmetric with respect to the axis $s=0$.

\vspace{4mm}
\begin{minipage}{8cm}
\includegraphics[width=8cm,height=6cm]{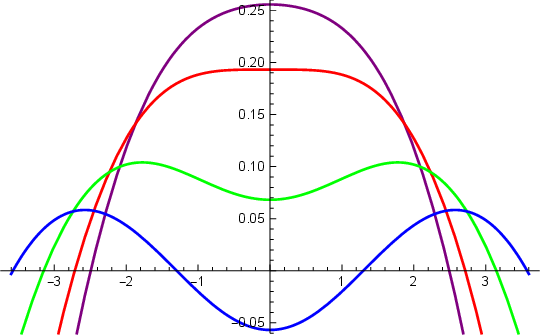}
\end{minipage}
\hfill
\begin{minipage}{8cm}
Fig. 4. Plots of the function $\hat E(\mu_c;s)$ from \eqref{BIS} at the same values of $p$ as in Fig. 3.
All curves are symmetric with respect to the $s=0$ axis as it should be
for the even function $\hat E(\mu_c;s)$.
\end{minipage}
\vspace{4mm}

The extremum positions of the function $E(\mu_c;y)$ at $\bar y_0=p/2$ seen in Fig. 3
are mapped to a single point $s=0$ in Fig. 4. The heights of these extrema are given by
\be\label{MIP}
\hat E(\mu_c;0)=\ln2-\frac p8\,.
\ee

At this point, it becomes interesting to expand the function $\hat E(\mu_c;s)$ in powers
of its argument $s$ around the origin. This yields
\be\label{LAE}
\hat E(\mu_c;s)\simeq\hat E(\mu_c;0)+\frac{p-4}{8p}\,s^2-\frac{s^4}{192}+O(s^6)
\ee
with the constant term $E(\mu_c;0)$ given in \eqref{MIP}.

The functional form appearing in \eqref{LAE} precisely coincides with that of the
usual Landau expansion with the critical value $p_c=4$ of the temperature parameter
$p$, at which the
$O(s^2)$ term changes its sign, while the stability of the underlying physical system is guaranteed by the correct sign of the $s^4$ term.
This actually implies that in the strong-repulsion limit $J_2\gg J_1$ and the
critical value of the chemical potential $\mu_c$,
the Curie-Weiss model defined in Section \ref{REX} belongs to the wide universality class given by the Landau expansion \eqref{LAE}. The closest in spirit representatives of this universality class are the lattice gas and other
Ising-like systems in zero external field.

In the following section, we shall consider the behavior of the basic function
$E(y)$ from \eqref{BIF} for chemical potentials $\mu$ differing from their
critical value $\mu_c$.

\subsection{Temperature dependence of $E(y)$ with $\mu\ne\mu_c$}

In the present section, we are going to consider a quite generic situation
where we relax the conditions $u(y)=0$ (see \eqref{SHN}),
$E'''(y)=0$ (see \eqref{POT}), and $\mu=\mu_c$ accepted above,
and look at special points where the first and the second derivatives of the function $E(y)$ vanish simultaneously. Recall that $E(y)$ is defined in \eqref{BIF} via
\be\label{BIA}
E(y)=\ln\left(1+\e^{y-p(\frac12f-\mu)}\right)-\frac{y^2}{2p}\,.
\ee

Its first derivative $E'(y)$ is given in \eqref{POE}. It becomes zero when $p\,x-y-yx=0$, and hence (cf. \eqref{EXC}),
\be\label{C1}
y=p\,\frac x{1+x}\,.
\ee
The second derivative $E''(y)$ from \eqref{POD} vanishes when $x^2+(2-p)\,x+1=0$,
and this quadratic equation has the solutions
\be\label{EEE}
x_{1,2}(p)=\frac p2-1\pm\sqrt{p\,\Big(\frac p4-1\Big)}\,.
\ee
When $p<p_c=4$, there are no real solutions $x_{1,2}$, and $E''(y)$
is always negative as it should be (see \eqref{POD}).
At $p=p_c=4$, the first real solution appears, $x_0=1$. This special case has been studied in Section \ref{SWY}.

And now, we are interested in the region $p>4$ where
we obtain two different solutions from \eqref{EEE}.
Vanishing of $E''(y)$ at $x=x_{1,2}(p)$ means that it happens when
(here we recall the definitions \eqref{SHN})
\be\label{YRT}
\e^{u(y)}=x_{1,2}(p),\mm{that is}u(y)=\ln x_{1,2}(p),\mm{and thus,}
y+p\,\mu-\frac12\,pf=\ln x_{1,2}(p)\,.
\ee

Threfore, $E'(y)$ and $E''(y)$ vanish simultaneously when we combine the conditions \eqref{C1}
and \eqref{EEE} into
\be\label{Y12}
y=\tilde y_{1,2}(p)=p\,\frac{x_{1,2}(p)}{1+x_{1,2}(p)}\,,
\ee
which defines the coordinates $\tilde y_1(p)$ and $\tilde y_2(p)$ of new special points.
Reporting \eqref{Y12} to the last equation in \eqref{YRT} leads us to the values of
$\mu$, with which the present setting is possible, namely
\be\label{M12}
\mu_{1,2}=\mu_{1,2}(p)=\frac12\,f-\frac{x_{1,2}(p)}{1+x_{1,2}(p)}+
\frac1p\,\ln x_{1,2}(p)\,.
\ee
For such $\mu_1$ and $\mu_2$ with any $p>4$, we have
$E'(\tilde y_{1,2})=E''(\tilde y_{1,2})=0$, and the function $E(y)$ has horizontal
inflection points with $E'''(\tilde y_{1,2})>0$ (see \eqref{POT}).

\subsubsection{An example calculation}

Let us do an example by calculating the corresponding curves at $p=6$.
This is the same choice of the temperature $p$ as in figures 3\,a) and 3\,b) in
\cite[p. 13]{KKD20}. At $p=6$, we have $x_{1,2}=2\pm\sqrt3$\,, the coordinates of inflection points are
$\displaystyle{\tilde y_{1,2}=6\,\frac{2\pm\sqrt3}{3\pm\sqrt3}}$\,, and the associated values of the chemical potential are
\be\label{MMM}
\mu_{1,2}=\frac12\,f-\frac{2\pm\sqrt3}{3\pm\sqrt3}+
\frac16\,\ln\!\big(2\pm\sqrt3\,\big)\,.
\ee
The plots of corresponding functions $E(y)$ with $f=10$ are given in Fig. 5.

\vspace{4mm}
\begin{minipage}{8cm}
\includegraphics[width=8cm,height=6cm]{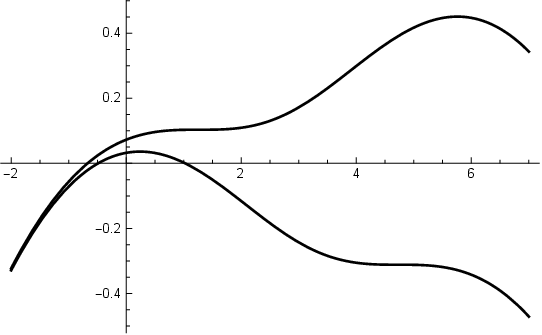}
\end{minipage}
\hfill\quad
\begin{minipage}{8cm}
Fig. 5. Plots of the function $E(y)$ from \eqref{BIF} with $p=6$, $f=10$, and
$\mu_1\simeq4.43<\mu_c$ (lower curve) and $\mu_2\simeq4.57>\mu_c$ (upper curve) given in \eqref{MMM} with plus and minus signs, respectively.
The respective ordinates of inflection points are approximately $-0.31$ and $0.10$.
\end{minipage}
\vspace{4mm}

Our Figure 5 combines, in fact, the plots of the same kind as in Figures 3\,a) and 3\,b) from \cite[p. 13]{KKD20}, into one. Indeed, they are very similar, though produced in a different setting involving the infinite sum over $n$ instead of strongly truncated one in \eqref{FUS}, and different physical parameters apart from $p=6$.

\subsubsection{A note on symmetry}

A numerical check shows that the inequality
\be
\mu_1<\mu_c<\mu_2\MM{with}\mu_c=\frac12(f-1)
\ee
holds for chemical potentials $\mu_1$ and $\mu_2$ from \eqref{MMM} related to inflection points of $E(y)$ at $p=6$. This suggests that it could be a good idea to write these values as
\begin{align}\label{BIV}
&
\mu_1=\mu_c+\frac12-\frac{2+\sqrt3}{3+\sqrt3}+\frac16\,\ln\!\big(2+\sqrt3\,\big),
\\\label{BIW}&
\mu_2=\mu_c+\frac12-\frac{2-\sqrt3}{3-\sqrt3}+\frac16\,\ln\!\big(2-\sqrt3\,\big)\,.
\end{align}
Let us define \emph{positive} deviations $\Delta\mu_1$ and $\Delta\mu_2$
of chemical potentials $\mu_1$ and $\mu_2$ from the central value $\mu_c$ via
\begin{align}\nn
&
\Delta\mu_1\equiv\mu_c-\mu_1=-\frac12+\frac{2+\sqrt3}{3+\sqrt3}-\frac16\,\ln\!\big(2+\sqrt3\,\big),
\\\nn&
\Delta\mu_2\equiv\mu_2-\mu_c=\frac12-\frac{2-\sqrt3}{3-\sqrt3}+\frac16\,\ln\!\big(2-\sqrt3\,\big)\,.
\end{align}
A straightforward algebraic calculation shows that
\be\nn
\Delta\mu_1=\Delta\mu_2\,,
\ee
and hence, $\mu_1+\mu_2=2\mu_c=f-1$.

The values $\mu_1$ and $\mu_2$ from \eqref{MMM} are symmetric with respect to the critical value of chemical potential, $\mu_c$ (see \eqref{MUC}).

In fact, the same symmetry property persists for generic values of $\mu_1(p)$
and $\mu_2(p)$ defined in \eqref{M12} for the range $p>4$. The difference of their distances to $\mu_c$ is given by
\be
\Delta\mu_1(p)-\Delta\mu_2(p)=-1+\frac{x_1(p)}{1+x_1(p)}+\frac{x_2(p)}{1+x_2(p)}-
\frac1p\,\ln\Big(x_1(p)\,x_2(p)\Big)=0\,.
\ee
This combination vanishes because $x_1(p)\,x_2(p)=1$, which can be easily inferred from the
definition of $x_{1,2}(p)$ in \eqref{EEE}. As in the special case considered just above, we have $\Delta\mu_1(p)=\Delta\mu_2(p)$ and $\mu_1(p)+\mu_2(p)=2\mu_c$ for any $p>4$.

\subsubsection{Consequences of changes in $\mu$ at fixed temperature $p>p_c$}

We conclude this section concerning the large-$f$ limit, by showing the behavior
of the function $E(y)$ (see \eqref{BIF} and \eqref{BIA}) and its maxima at the fixed temperature, $p=6$, and varying chemical potential $\mu$. The plots are given in Figure 6.

The first, blue curve, corresponds to the lowest value of $\mu$, $\mu=4<\mu_c$, while $\mu=4.5$.
The red curve is drawn at $\mu=\mu_c$, the same curve is shown in blue in Fig. 3.
The two black curves with inflection points are the same as in Fig. 5.
The magenta curve corresponds to the largest value of $\mu$, $\mu=4.64$.

\vspace{4mm}
\begin{minipage}{8cm}
\includegraphics[width=9cm,height=6cm]{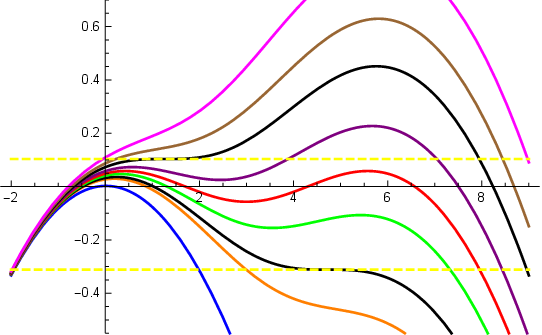}
\end{minipage}
\hfill\quad
\begin{minipage}{7cm}
Fig. 6. Plots of the function $E(y)$ from \eqref{BIF} with $p=6$, $f=10$, and
growing values of $\mu$: 4, 4.4, $\mu_1$ from \eqref{MMM} and \eqref{BIV}, 4.47,
$\mu_c=4.5$, 4.53, $\mu_2$ from \eqref{MMM} and \eqref{BIW}, 4.6, 4.64.
The yellow dotted lines indicate the values of $E(y)$ at the inflection points
$\tilde y_1$ and $\tilde y_2$ given explicitly in the line preceding \eqref{MMM}.
\end{minipage}
\vspace{4mm}

By the Laplace method, in the limit $N\to\infty$, the value of the integral over $y$ in \eqref{BIF} is determined by the height of the global maximum attained by the function $E(y)$. If $\mu<\mu_c$, such global maxima are situated
roughly between $y=0$ and $y=1$; see blue, orange, black, and green curves in the
bottom of Fig. 6.

The red curve at $\mu=\mu_c$ signals a first-order phase transition.
As soon as $\mu$ exceeds $\mu_c$, the global maximum jumps to a position
between $y=5$ and $6$. The "new" family of global maxima for $\mu>\mu_c$ is
shown by the purple, (second) black, brown, and magenta curves.

\vspace{2mm}
Finally, let us sketch the last consequence of $\mu$ being not equal to $\mu_c$.

By analogy with \eqref{BIS}, let us shift in \eqref{BIF} or \eqref{BIA} the integration variable $y$ via $y=s+pf/2-p\mu$. Thus we obtain
\be\label{MIS}
\hat E(\mu;s)=\ln\left(1+\e^s\right)
-\frac1{2p}\,\Big(s+p\Big(\frac12+\mu_c-\mu\Big)\Big)^2,
\ee
where we took into into account the definition of $\mu_c$ in \eqref{MUC}.
Expanding the last function in powers of $s$ we obtain, similarly as in \eqref{LAE},
\be\label{MIC}
\hat E(\mu;s)=\ln2-\frac p2\Big(\mu-\mu_c-\frac12\Big)^2
+(\mu-\mu_c)\,s+\,\frac{p-4}{8p}\,s^2-\frac{s^4}{192}+O(s^6).
\ee
At $\mu=\mu_c$, the last expansion becomes an even function and reduces to the one in \eqref{LAE}.

Similarly as at the end of Section \ref{SPE}, we could speculate here about certain equivalence of the present model with large enough repulsion between
particles within a cell and such systems as lattice gas with non-zero
chemical potential or Ising systems in the presence of an external ordering
field.

\section{Summary and outlook}

On the physical side of our paper, we propose the first analytical calculations for certain special cases of the Curie-Weiss cell model of fluid introduced in \cite{KKD20}
and described in Section \ref{REX}.

First of all, explicit results for the simplest special cases of the ideal gas and the high-temperature limit are derived in agreement with the well-known classical data (see Section \ref{SSS}).

Moreover, we have succeeded to show that the marginal case of equal attraction and repulsion interactions between the particles ($J_1=J_2$ in \eqref{CV}) is meaningful.
This limitation does not lead to any unphysical divergence of the integral representing the grand-canonical partition function of the system (see Sections \ref{SIS} and \ref{SII}). An explicit calculation in the asymptotic regime $z\to+\infty$ showed a relation of this marginal equal-interaction case to the ideal-gas limit.

We have performed an extended analytical study of the strong-repulsion limit
$J_2\gg J_1$ in Section \ref{J2B} supported by numerous graphical representations
and accompanied by discussions of related physical implications.

\vspace{2mm}
Explicit calculations performed in Sections \ref{SIS} and \ref{SII} and their physical conclusions would be impossible without certain progress on the mathematical side.

\vspace{1mm}
Thus, for the announced in the Abstract discrete Gauss-Poisson probability distribution function \eqref{GPD}, we have found the asymptotic behavior of its normalization $R(r;z)$, given by the infinite sum \eqref{R},
as $z\to+\infty$. In Section \ref{RRR} we recorded the related results without proof with a detailed exposition of the derivation planned for the nearest future.

The asymptotic formulas found for the function $\ln R(r;z)$ 
provide quite accurate approximations in a wide range of $z$ (down to $z\approx3$) which is illustrated on the left sections of Figures 1 and 2.

At large enough values of $r$ exceeding some threshold value $r^*$, the difference \eqref{DIF} between the function
$r\ln R(r;z)$ and its asymptotics becomes oscillatory. This interesting behavior is illustrated on the right-hand side of Figure 2.
At this point, we would like to mention two other instances \cite[Sec. 4]{GRM17}, \cite[p. 11, Fig. 4]{Sh24},
where the periodic oscillatory behavior emerges in different situations interesting from the mathematical point of view.

This special kind of oscillatory behavior certainly deserves a further investigation.

\section*{Acknowledgements}

We are grateful to our colleagues in the NRFU project,
M.P. Kozlovskii, I.V. Pylyuk, and R.V. Romanik,
for weekly enlightening and motivating discussions.
A careful reading of the draft by R.V. Romanik and his suggestions are gratefully acknowledged. Special thanks of MAS are to M.P. Kozlovskii for his personal invitation to join the project.

The financial support by the National Research Foundation of Ukraine under the project No. 2023.03/0201 is gratefully acknowledged.

\addcontentsline{toc}{section}{References}
\providecommand{\href}[2]{#2}
\end{document}